\input amstex
\documentstyle{amsppt}
{\catcode`\@=11\gdef\logo@{}} \footnote"" {Supported by grant
VEGA1/7146/20 } \topmatter
\title
Principle conditioning
\endtitle
\author
       Olga N\'an\'asiov\'a,
\endauthor
\affil
      Slovak Technical University
\endaffil
\date
November 2001
\enddate
\address
       Department of Mathematics and Descriptive Geometry
\newline
       Faculty of Civil  Engineering STU
\newline
    Radlinsk\'eho  11
\newline
    813 68 Bratislava
\newline
    Slovak  Republic
\endaddress
\email
     olga\@vox.svf.stuba.sk
\endemail
\abstract The definition of the  conditional probability is very important
in the theory of
  the probability. This definition is
 based on the
fact, that random events can be simultaneously measurable. This paper
 deal with the problem of conditioning for such random events, which
 are not simultaneously measurable.
 This paper defines conditional
states as convex combination of special states.

\endabstract
\endtopmatter
\document
\head
       {\bf  Introduction}
\endhead
\vskip 2pc
    The classical Kolmogorovian model for random events was developed
only for such random events, which are simultaneously measurable
 (in another words, which are compatible). The basic algebraic structure,
which is used as a model  for non-compatible random events is an
orthomodular lattice (OML), or an orthomodular $\sigma$-lattice
($\sigma $-OML). In this paper we  determine a conditional
    state (analogical notion of conditional probability) as
    a convex combination of ``orthogonal states `` on  an OML.

     In the classical theory we assume that  random
    events can be interpreted as a set of outcomes of experiments. A
    probability space is a triple ([1],[2],[8])
$(\Omega, \Cal B ,P)$, where
$\Omega $
    is a set of all elementary random events, $\Cal B$ is a
$\sigma $-algebra
    of subset of
$\Omega $ and $P$
    is a probability
    measure. In the non-commutative approach we have a couple
$(L,M)$,
    where
$L$
    is a
$\sigma$-OML
    and
$M$
    is a set of states on it.

    Let
$(\Omega_i ,\Cal F_i)$ for $i=1,...,n$
    be  measurable spaces. Let
$ \overline{\Omega }=\Omega_1\times ...\times\Omega_n  $.
    If
 $\overline{\omega }=(\omega_1,...,\omega_n)$,
    then
$\pi_i(\overline{\omega})
 =\omega_i$. Then $L=\{\pi_i^{-1}(A);\quad A\in\Cal F,\quad i=1,...,n\}$,
where for example $\pi_1^{-1}(A)=(A,\Omega,...,\Omega)$, for
$A\in\Cal F_1$. Then $L$ can be organized as an OML by the
following way: \roster \item $\pi_i^{-1}(\Omega ):=1$ ; \item
$\pi_i^{-1}(A)\vee\pi_i^{-1}(B):=\pi_i^{-1}(A\cup B)$ and
 $\pi_i^{-1}(A)\vee\pi_j^{-1}(B):=1$, for $i\neq j$;
\item  $\pi_i^{-1}(\emptyset ):=0$;
\item $\pi_i^{-1}(A)\perp \pi_j^{-1}(B)$ if and only if $i=j$ and
$A\cap B=\emptyset $.
\endroster
\   Let
$(\Omega_i ,\Cal F_i,P_i)$ for $i=1,...,n$
    be some probability spaces and
$L$
    be the OML defined as before. A map
$$m:L\to [0,1],$$
    such that
$$m(\pi_i^{-1}(A))=P_i(A)\quad\quad \text
{for each }\quad\quad A\in\Cal F_i $$
    is a state on
$L$.

 Well known examples of OMLs include Boolean algebras and the
 orthogonal projections on a Hilbert space.
\definition {Definition 1 } [3] Let $L$ be a nonempty set
    endowed with a partial order $\leq $ with the
    largest element ($1$) and the smallest element ($0$). Let there
    be defined the operations of supremum ($\vee $), infimum ($\wedge $)
    (the lattice operations ) and a map $\perp :L\to L$ with the
    following properties:
\roster \item "\it (i)"  For any $\{a_n\}_{n\in\Cal A}\in L$,
where $\Cal A\subset N$ is finite ($\Cal A$ is countable)
$$\bigvee_{n\in\Cal A} a_n,
                 \bigwedge_{n\in\Cal A} a_n\in L\text {.}$$
\item "\it (ii)" For any $a\in L$ $(a^\bot )^\bot =a$.
\item "\it (iii)" If $a\in L$,  then $a\vee a^\bot =1$.
\item "\it (iv)"  If $a,b\in L$ such that $a\le b$,  then $b^\bot\le a^\bot $.
\item "\it (v)"  If $a,b\in L$  such that $a\le b$ then $b=a\vee (a^\bot\wedge b)$ (orthomodular law).
\endroster
    Then
$(L,0,1,\vee ,\wedge ,\perp )$
    is called {\it an orthomodular lattice} (briefly
{\it $L$ is an OML}) ({\it a $\sigma $-OML}.
\enddefinition
\vskip 1pc

\vskip 1pc
     Let $L$ be an  OML ( a $\sigma $-OML). Then the elements
$a,b\in L$
    will be called:
\roster
\item {\it orthogonal} ($a\bot b$) iff $a\le b^\bot$;
\item {\it compatible} ($a\leftrightarrow b$) iff  there exist mutually
    orthogonal elements $a_1,b_1,c\in L$   such that
    $$a=a_1\vee c\quad \text {and}\quad  b=b_1\vee c\text {.}$$
\endroster
 If  $a_i\in L$ for any $i\in\Cal A$ and
$b\in L$ is such that $b\leftrightarrow a_i $ for all $i$, then $
b\leftrightarrow\bigvee_{i\in\Cal A} a_i$ and
$$b\wedge \bigvee_{i\in\Cal A} a_i=\bigvee_{i\in\Cal A} a_i\wedge b$$
([10]).
\vskip 1pc
\definition {Definition 2 }[3]
 A map $m:L\to R $  such that
\roster\item "\it (i)"
    $m(0)=0$ and $m(1)=1$,
\item "\it (ii)"
    if $a\bot b$ then $m(a\vee b)=m(a)+m(b)$,
\endroster
    is called {\it a state} on $L$. If $L$ is
   a $\sigma $-OML and $m$ is
   a $\sigma $-additive function then $m$ will be called a
    {\it  $\sigma $-state}.
\enddefinition
\vskip 3pc \head {\bf 1.  A conditional state on an OML}
\endhead
\vskip 2pc
\definition {Definition 1.1}
    Let $L$ be an OML. A subset
$L_0\subset L-\{0\} $ is
    called {\it a conditional system (a CS)   {\it ( a $\sigma$-CS
    }) if the following conditions are fulfilled:
\roster \item If $a,b\in L_0$, then $a\vee b\in L_0$. (If $a_i\in
L_0$, for
    $i=1,2,...$, then $\bigvee_i a_i\in L_0$.)
\item If $a,b\in L_0$ and $a < b$, then $a^\perp\wedge b\in L_0$.
\endroster
\enddefinition
\definition {Definition 1.2}
    Let $L$ be an OML and $L_0$ be a CS ( a $\sigma $-CS).
    Let $$f:L\times L_0\to [0,1]\text {.}$$
 If the function $f$ fulfils the
    following conditions:
\roster \item "(C1)" for each $a\in L_0$ $f(.,a)$ is a state on
$L$ ($\sigma $-state); \item  "(C2)" for each $a\in L_0$
$f(a,a)=1$; \item "(C3)" if $\{a_i\}_{i\in\Cal A}\in L_0$, where
$\Cal A\subset N$, $\Cal A$ has finite (countable) cardinality and
    $a_i$ are mutually orthogonal, then for each $b\in L$
    $$f(b,\bigvee_{i\in\Cal A}a_i)=\sum_{i\in\Cal A}f(a_i,
    \bigvee_{i\in\Cal A}a_i)
    f(b,a_i);$$
\endroster
    then $f$ is called {\it a conditional state}
({\it $\sigma $-conditional state)}.
\enddefinition
\vskip 2pc It is clear, that if  $L$ is a $\sigma $-OML,
$\{a_i\}_{i\in\Cal A}$, where $\Cal A\subset N$, such that
$a_i\perp a_j$, for $i\ne j$, than we can rewrite the Proposition
1.1 for a $\sigma  $-conditional state. Moreover for any
$\{a_i\}_{i\in\Cal A}$ there exists many conditional states (or
$\sigma $-conditional states). On the other hand, because a
measurable space can be described  as a $\sigma $-OML [3], then
this representation is fulfilled  for a probability space, too.

\vskip 1pc
It is clear, that if there  exists a probability
measure $\mu $ on the measurable space  $(\Omega, \Cal B)$ ,
then the  conditional probability $f$ exists on
$\Cal B\times\Cal B_0 $ and
 $$f(A,B)=\frac{\mu (A\cap B)}{\mu (B)}{\text ,}$$
 where $\Cal B_0\subset \{E\in \Cal B;\quad
\mu (E)\neq 0\}$. The system $(\Omega ,\Cal B ,\Cal B_0, f)$ is
called the conditional probability system (CPS).

Let $\Cal P$ be some collection
 of probability measures on
$(\Omega, \Cal B)$.
 It is a question, when this collection $\Cal
P$ can be organized as a system of conditional probabilities. On
the classical theory of probability  the following theorems are
fulfilled: \proclaim {Proposition 1.1} Let $(\Omega ,\Cal B, \Cal
B_0, f)$ be a CPS.  Let
 $\{B_i\}_{i\in\Cal A}\in\Cal B_0$,
$\Cal A\subset N$ and  let there exist $B\in\Cal B_0$, such that
$f(B,B_i)=1$ and $f(B_i,B)>0$ for any $i\in \Cal A$. Then, for
each $C\in\Cal B$
$$f(C,B)=\sum_{i\in\Cal A}f(C,B_i)f(B_i,B)$$
iff $$  f(\bigcup_{i\in\Cal A}B_i,B)=\sum_{i\in\Cal A}f(B_i,B)=1{\text .}$$
\endproclaim
\vskip 1pc \proclaim {Proposition 1.2} Let $(\Omega ,\Cal B, \Cal
B_0, f)$ be  a CPS.  Let
 $\{B_i\}_{i\in\Cal A}\in\Cal B_0$,
$\Cal A\subset N$ and  let there exist $B\in\Cal B_0$, such that
$f(B,B_i)=1$ and $f(B_i,B)>0$ for any $i\in \Cal A$. Then, for
each $C\in\Cal B$
$$f(C,B)=\sum_{i\in\Cal A}f(C,B_i)f(B_i,B)$$
and for any $i\neq j$ $f(B_i,B_j)=0$.
\endproclaim
\vskip 1pc From this approach follows, that the definition of a
conditional state ($\sigma $-conditio\-nal state) on an OML (a
$\sigma $-OML) has been defined correctly. More details about the
classical approach to the conditional probability we can find for
example in [9].
\vskip 2pc
\proclaim {Proposition 1.3 }
    Let $L$ be an OML. Let $\{a_i\}_{i=1}^n\in L$, $n\in N$
    where  $a_i\perp a_j$ for $i\neq j$. Let
    for any $i$ there exist a state $\alpha_i$,
    such that $\alpha_i(a_i)=1$. Then there
exists a CS such that for any
    ${\bold k}=(k_1,k_2,...,k_n)$, where $k_i\in [0;1]$ for
    $i\in\{1,2,...,n\}$ with the property
    $\sum_{i=1}^n k_i=1$
     there exists a conditional state
    $$f_{\bold k}:L\times L_0\to [0,1],$$
    and
\roster
\item
    for any $i$ and each $d\in L$
    $f_{\bold k}(d,a_i)=\alpha_i(d);$
\item
    for each $a_i$
    $$f_{\bold k}(a_i,\bigvee_{j=1}^na_j)=k_i\text {;}$$
\endroster
\endproclaim
\demo {Proof }
    Let
$$L_0=\{c\in L;\quad c=\bigvee_{j\in\Cal A}a_j,
\quad\text {for each}\quad \Cal A\subset \{1,2,...,n\}\}\text {.}$$
    Then it is clear that
    $L_0$ is a CS and so $L_0$ exists in $L$.

    From the assumption, we have the set of triples
    $\{(\alpha_i,a_i,k_i),i=1,...,n\}$ and
    from the properties of
    a CS
    follows that for each
    $c\in L_0$
    there exist
    $\{i_1,...,i_s\}\subset \{1,...,n\},$ such that
    $$c=\bigvee_{j=1}^s a_{i_j}\quad \text
    {and}\quad
    \alpha_{i_j}(a_{i_j})=1.$$
    Let as denote
    $\Cal K(c)=\sum_{j=1}^sk_{i_j}$.

    Let $f_{\bold k}:L\times L_0\to [0,1]$
    such that for each $d\in L$ and $c\in L_0$
    $$f_{\bold k}(d,c)=\frac{1}{\Cal K(c)}
    \sum_{j=1}^sk_{i_j}\alpha_{i_j}(d).$$

    Now we show, that $f_{\bold k}$ is the  conditional state.

    (C1) Let $c\in L_0$. Then
$$f_{\bold k}(1,c)=\frac{1}{\Cal K(c)}\sum_{j=1}^sk_{i_j}\alpha_{i_j}(1)
    =\frac{1}{\Cal K(c)}\sum_jk_{i_j}=
    \frac{\Cal K(c)}{\Cal K(c)}=1 $$
    and
$$f_{\bold k}(0,c)=\frac{1}{\Cal K(c)}\sum_jk_{i_j}\alpha_{i_j}(0)
    =\frac{1}{\Cal K(c)}\sum_jk_{i_j}.0=
    \frac{0}{\Cal K(c)}=0. $$

    Let $d,b\in L$, such that $
    d\perp b$. Then
$$\align
     f_{\bold k}(d\vee b,c)&=\frac{1}{\Cal K(c)}
    \sum_jk_{i_j}\alpha_{i_j}(d\vee b)
    =\frac{1}{\Cal K(c)}(\sum_jk_{i_j}\alpha_{i_j}(d)+
    \sum_jk_{i_j}\alpha_j(b))\\
    &=\frac{1}{K(c)}\sum_jk_{i_j}\alpha_{i_j}(d)+
    \frac{1}{ K(c)}\sum_jk_{i_j}\alpha_{i_j}(b)
    =f_{\bold k}(d,c)+f_{\bold k}(b,c).
\endalign $$

    So $f_{\bold k}$ is a state on $L$.

(C2)
    It is easy to see that, for each $c\in L_0$
$$f_{\bold k}(c,c)=1.$$

(C3) It is enough to show it for two orthogonal elements from
$L_0$. Let
    $c_1,c_2$ be such elements from $L_0$, that
    $$c_1=\bigvee_{i=1}^{n_1}a_i\quad\text { and}\quad
    c_2=\bigvee_{i=n_1+1}^{n_2}a_i.$$
    Then
$$f_{\bold k}(c_j,c_1\vee c_2)=\frac{\Cal K(c_j)}{\Cal K(c_1\vee c_2)}
    \quad \text {for}\quad j=1,2$$

    and
$$f_{\bold k}(d,c_1\vee c_2)=\frac{1}{\Cal K(c_1\vee c_2)}
\sum_{i=1}^{n_2}k_i\alpha_i(d).$$

From it follows that
$$\align &\sum_{j=1}^{2}f_{\bold k}(c_j,c_1\vee c_2)
f_{\bold k}(d,c_j)=\\
& =\frac{\Cal K(c_1)}{\Cal K(c_1\vee c_2)}\frac{1}{\Cal K(c_1)}
\sum_{i=1}^{n_1}k_i\alpha_i(d)+\frac{\Cal K(c_2)}{\Cal K(c_1\vee
c_2)}\frac{1}{\Cal K(c_2)} \sum_{i=n_1+1}^{n_2}k_i\alpha_i(d)
\\&=\frac{1}{\Cal K(c_1\vee
c_2)}\sum_{i=1}^{n_2}k_i\alpha_i(d)=f_{\bold k}(d,c_1\vee
c_2).\endalign
$$
So $f_{\bold k}$ is the conditional state.

Let $a=\bigvee_{i=1}^na_i$. Then
$$f(.,a)=\frac{1}{\Cal K(a)}\sum_ik_i\alpha_i(.)=\sum_ik_i\alpha_i(.),$$
and then  for each $i=1,2,...,n$
$$f(a_i,a)=k_i.$$
From it follows, that for each $d\in L$ and $a_i$ $i=1,2,...,n $

$$f_{\bold k}(d,a_i)=\frac{1}{\Cal K(a_i)}k_i\alpha_i(d)=\alpha_i(d).$$
It is clear that $f_{\bold k}:L\times L_0\to [0,1]$ is the
conditional state with the properties (1) and (2).

 (Q.E.D.)
\enddemo

\vskip 3pc
\head {\bf 2. Dependence and  independence} \endhead
\vskip 3pc
\definition {Definition 2.1}
     Let $L$ be an OML and $f$ be a  conditional state.
     Let $b\in L$, $a,c\in L_0$ such that $f(c,a)=1$. Then $b$ is
 independent
      of $a$, with respect to the state $f(.,c)$ ($b\asymp_{f(.,c)}a$)
      iff $f(b,a)=f(b,c)$.
\enddefinition
 The classical   definition of independence  a probability space
$(\Omega,\Cal B,P)$ is a special case of this definition because
$$P(B|A)=P(B|\Omega )
\quad \text {iff}\quad P(B\cap A|\Omega )=P(B|\Omega )P(A|\Omega
).$$
     Let $L$ is a OML. Let $a_1,...,a_n\in L-\{0\}$, such that
    $a_i\perp a_j$, for $i\neq j$. Let $\alpha_i$ $i=1,...,n$ be
    a state such that
    $\alpha_i(a_j)=\delta_{i,j}$, the Kroneker
 $\delta_{i,j} $ which $=1$ when   $i=j$ and $0$ otherwise.
    Then, for each $k_i\in [0,1]$ (i=1,...,n), such that
 $$\sum_{i=1}^nk_i=1\quad \text {a map }\quad\mu :=
 \sum_{i=1}^nk_i\alpha_i=f_\mu(.,\bigvee_{i=1}^n a_i) $$
 is a state and we say that $\alpha_i$ is a conditional state
 with the condition $a_i$
 ($\alpha_i=f_\mu (., a_i))$ and $k_i=\mu
 (a_i)$. Then for $b\in L$
 $$b\asymp_{f_\mu }a_i\quad\text { iff }\quad\alpha_i(b)
 =\mu (b).$$
\proclaim {Proposition 2.1}
 Let $L$ be a OML. Let $a_1,...,a_n\in L-\{0\}$, such that
 $a_i\perp a_j$, for $i\neq j$. Let $\alpha_i$ $i=1,...,n$ be
 a state such that
 $\alpha_i(a_j)=\delta_{i,j}$.
 Let $k_i\in [0,1]$ (i=1,...,n), such that
 $$\sum_{i=1}^nk_i=1\quad \text {and let }\quad\mu =
 \sum_{i=1}^nk_i\alpha_i .$$
 Then
 \roster\item
  $b\asymp_{f_\mu }a_i$, iff $b\asymp_{f_\mu}\vee_{j\neq i}a_j$;
 \item
  $b\asymp_{f_\mu }a_i$, iff $b^\perp\asymp_{f_\mu} a_i$.
 \endroster
 \endproclaim
\demo {Proof} \roster\item It is enough to show it for $i=1$. Let
$b\asymp_{f_\mu }a_1$, then it follows that $\alpha_1(b)=\mu(b)$
and so
$$\align\mu (b)=f_\mu(b,\bigvee_{i=1}^n
a_i)=&f_\mu(a_1,\bigvee_{j=2}^na_j)f_\mu(b,a_1)+f_\mu(\bigvee_{j=2}^na_j,
\bigvee_{i=1}^na_i) f_\mu(b,\bigvee_{j=2}^na_j)
\\
\alpha_1(b)=&k_1\alpha_1(b)+f_\mu(\bigvee_{j=2}^na_j,
\bigvee_{i=1}^na_i) f_\mu(b,\bigvee_{j=2}^na_j)\\
(1-k_1)\alpha_1(b)=&f_\mu(\bigvee_{j=2}^na_j, \bigvee_{i=1}^na_i)
f_\mu(b,\bigvee_{j=2}^na_j)\\
f_\mu(\bigvee_{j=2}^na_j, \bigvee_{i=1}^na_i)\alpha_1(b)=&
f_\mu(\bigvee_{j=2}^na_j, \bigvee_{i=1}^na_i)
f_\mu(b,\bigvee_{j=2}^na_j)\\
\alpha_1(b)=&f_\mu(b,a_1)=f_\mu(b,\bigvee_{j=2}^na_j) .\endalign
$$ From it follows that $b\asymp_{f_\mu}\vee_{j\neq i}a_j$. The
converse implication can be shown analogously.

 \item If $b\asymp_{\mu
}a_i$, then
$$\mu (b^\perp )=1-\mu (b)=1-\alpha_i(b)=\alpha_i(b^\perp )$$
and so $b^\perp\asymp_{\mu }a_i$. The converse  implication can be
shown analogously.
 \endroster
 (Q.E.D.)
 \enddemo
 \proclaim {Proposition 2.2}
 Let $L$ be an OML, $L_0$ be  a CS and  $f:L\times L_0\to [0,1]$
 be a conditional state.
 \roster\item
 Let $a^\perp,a,c\in L_0$, $b\in L$ and $f(c,a)=f(c,a^\perp )=1$. Then
 $b\asymp_{f(.,c)} a$  iff $b\asymp_{f(.,c)} a^\perp$.
 \item
  Let $a,c\in L_0$, $b\in L$ and $f(c,a)=1$. Then
 $b\asymp_{f(.,c)} a$  iff $b^\perp \asymp_{f(.,c)} a$.
 \item
 Let $a,b,c\in L_0$, $b\leftrightarrow a$ and $f(c,a)=f(c,b)=1$,
 $f(a,b)\neq 0$, $f(b,a)\neq 0$. Then
 $b\asymp_{f(.,c)} a$  iff $a\asymp_{f(.,c)} b$.
 \item
 Let $b,c,d\in L_0$, $b\perp d$, $a\in L$ and $f(c,b)=f(c,d)=1$.
 If
 $a\asymp_{f(.,c)} b$ and $a\asymp_{f(.,c)} d$
 then $a\asymp_{f(.,c)}b\vee d$
 \endroster
 \endproclaim
\demo {Proof}

(1)  From the definition of a conditional state it follows that
for each $x\in L$
  $$f(x,c)=f(a,c)f(x,a)+f(a^\perp ,c)f(x,a^\perp ).\quad \quad (i)$$
  Let $b\asymp_{f(.,c)} a$. It means, that  $f(b,c)=f(b,a)$. If we put
  $x=b$, then  we get
   $$f(b,a)=f(b,c)=f(a,c)f(b,a)+f(a^\perp ,c)f(b,a^\perp ).$$
Thus
   $$(1-f(a,c))f(b,a)= f(a^\perp ,c)f(b,a^\perp ),$$
but $1-f(a,c)=f(a^\perp ,c).$ Then $$f(a^\perp ,c)f(b,a)=f(a^\perp
,c)f(b,a^\perp )$$ and so
   $$f(b,a)=f(b,a^\perp )=f(b,c).$$
Thus $b\asymp_{f(.,c)}a$.
   The converse  implication can be shown analogously.

(2)  Let $b\asymp_{f(.,c)} a$. Then $f(b,c)=f(b,a),$ and so
$1-f(b,c)=1-f(b,a).$
 Thus
$f(b^\perp ,c)=f(b^\perp ,a).$
  The converse  implication can be shown analogously.

(3) By (i) with $x=a\wedge b$, we have
 $$f(b,a)=f(a\wedge b,a).\quad\quad (ii)$$
On the other hand, by (i) with $b$ in place $a$, we have
$$f(a,b)=f(a\wedge b,b).$$
 From the definition of a conditional state, we can write
 $$f(x,c)=f(a,c)f(x,a)+f(a^\perp ,c)f(x,a^\perp ), $$
 for each $x\in L$.
If we put $x=a\wedge b$, then
 $$f(a\wedge b,c)=f(a,c)f(a\wedge b,a)=f(a,c)f(b,a).$$
On the other hand
$$f(x,c)=f(b,c)f(x,b)+f(b^\perp ,c)f(x,b^\perp ), $$
and  we get
$$f(a\wedge b,c)=f(b,c)f(a\wedge b,b)=f(b,c)f(a,b).$$
But $b\asymp_{f(.,c)} a$ so  $f(b,c)=f(b,a)$. Then
 $$f(a\wedge b,c)=f(b,c)f(a,b)=f(b,a)f(a,b)$$
 analogously
$$ f(a\wedge b,c)=f(a,c)f(b,a)$$
and, by (ii) we can write
 $$f(b,a)f(a,b)=f(a,c)f(b,a),$$
 so that, since $f(b,a)\neq 0$
 $$f(a,b)=f(a,c)$$  and   $a\asymp_{f(.,c)} b.$
  The converse  implication can be shown analogously.

(4) Let $b\perp d$, $f(c,b)=f(c,d)=1$.
 Then
 $$\align f(c,b\vee d)&=f(d,d\vee b)f(c,d)+f(b,d\vee b)f(c,b)\\
 &=f(d,d\vee b)+f(b,d\vee b)\\
 &=f(b\vee d,b\vee d)=1.\endalign $$
 If  $a\asymp_{f(.,c)} b$, $a\asymp_{f(.,c)} d$, then
 $f(a,b)=f(a,c)=f(a,d)$ and
 $$\align
f(a,b\vee d)&=f(d,d\vee b)f(a,d)+f(b,d\vee b)f(a,b)\\
 &=f(d,d\vee b)f(a,c)+f(b,d\vee b)f(a,c)\\
 &=f(b\vee d,b\vee d)f(a,c)=f(a,c).
\endalign $$

It means  $a\asymp_{f(.,c)}b\vee d$.

 (Q.E.D.)
\enddemo
\example {Example } Let $L=\{a,a^\perp ,b,b^\perp ,0,1\}$ and
$L_0=L-\{0\}$. Let $\alpha ,\alpha^,$ be states on $L$ such that
$\alpha (a)=\alpha^,(a^\perp )=1$ and let $\bold k=(0.1,0.9)$.
Then we can define a conditional state by the following way:
$$f_{\bold k}(d,a)=\alpha (d)\quad\text {and}
\quad f_{\bold k}(d,a^\perp )=\alpha^, ( d)$$
$$\align f_{\bold k}(d,1)&=0.1\alpha (d)+0.9\alpha^,(d)\\
 &=f_{\bold k}(b,1)f_{\bold k}(d,b)+f_{\bold k}(b^\perp,1)f_{\bold k}(d,b^\perp)
\endalign $$
for each $d\in L$.  Let $\alpha (b)=0.2$ and $\alpha^,(b)=0.3$.
Then
$f_{\bold k}(b,1)=0.29$   and we can write
 $$f_{\bold k}(d,1)=0.29f_{\bold k}(d,b)+0.71f_{\bold k}(d,b^\perp ).$$
 If we put $d=a$, then
 $$f_{\bold k}(a,b)\in [0,\frac{10}{29}]\quad\text {and}\quad
 f_{\bold k}(a,b^\perp)\in [0,\frac {10}{71}].$$
 Therefore
$$0.29=f_{\bold k}(b,1)\ne f_{\bold k}(b,a)=0.2\text {,}$$
 then
 $$b\quad\text { is not independent of}\quad a\quad\text {with respect to
the state}\quad f_{\bold k}(.,1)\text {.}$$
If  $f_{\bold k}(a,1)=0.1$, then $f_{\bold k}(a,1)=f_{\bold k}(a,b)$ and so
$$ a\quad\text {is  independent of}\quad b\quad\text
{with respect to the state}\quad f_{\bold k}(.,1) \quad
(a\asymp_{f_{\bold k}(.,1)} b)\text {.}$$ From the above mentioned
it follows that the Boolean algebra as a measurable system 
$B_1=\{0,1,a,a^\perp\}$ is independent of the Boolean algebra as a
measurable system
 $B_2=\{0,1,b,b^\perp \}$ with
respect to the conditional state $f_{\bold k}$, and $B_2$
 is dependent on the $B_1$ with respect to
the conditional state $f_{\bold k}$. It may be that this
approach to the conditional state can help   describe some
problems of causality in the theory of
 probability.

 \endexample
  

\Refs
\ref  \no 1\by Renyi A.  \paper On a new axiomatic theory of probability
\jour Acta Math. Acad.
    Sci. Hung.\vol 6 \yr 1955\pages 285-335
\endref

\ref  \no 2\by Renyi A. \paper  On conditional probabilities spaces generated by a
    dimensionally ordered set of measures\jour Teorija verojatnostej i jejo primenenija\vol 1
     \yr 1947\pages 930-948
\endref

\ref   \no 3\by Varadarajan V. \paper Geometry of quantum theory \jour Princeton,
    New Jersey, D. Van Nostrand,  \yr 1968
\endref

\ref \no 4\by  N\'an\'asiov\'a O.\paper A note on the independent events on quantum
logics\jour  Busefal\vol 76\yr 1998\pages  53-57
\endref
\ref   \no 5\by N\' an\' asiov\' a O. \paper On conditional probabilities on quantum logic.
     \jour  Int. Jour. of Theor. Phys. \vol 25 \yr 1987 \pages 155-162
\endref
\ref\no 6 \by  N\'an\'asiov\'a\paper Representation of conditional probability on a quantum logic.\jour Soft Computing\yr 2000
\endref
\ref \no 7\by Greechie D.\paper Orthomodular lattice admitting no states
\jour Jour. Cominat. Theory \vol 10\yr 1971\pages 119-132
\endref
\ref \no 8\by Kolmogoroff A.N.\paper Grund\-begriffe der
    Wahr\-scheikch\-keits\-rechnung      \jour Springer, Berlin\yr 1933
\endref
\ref \no 9 \by Rie\v can B.,Neubrun T.\paper Integral,  Measure, and Ordering
\jour Kluwer Acad. Publ., Dotrecht, Ister Science, Bratislava \yr 1997 \endref
\ref \no 10 \by Dvure\v censkij A., Pulmannov\'a S.\paper New Trends in Quantum Structures
\jour Kluwer Acad. Publ., Dortrecht, Ister Science, Bratislava
 \yr 2000  \endref
\ref\no 11\by Rie\v canov\' a Z.\paper Genera\-lization  of blocks for D-la\-ttice
and lattice ordered effect algebras\jour Pre\-print FEI STU, Bratislava \yr 1998 \endref
\ref\no 12\by Greechie R.J., Foulis D.J., Pulmannov\' a\paper The center of an effect algebra
\jour Order\vol 12\yr 1995\pages 91-106\endref
\ref \no 13 \by Gudder S.\paper Stochastic Methods in Quantum Mechanics
\jour Elsevier, North Holland\yr
1979\endref

\endRefs
\enddocument